\documentclass{ws-ijmpd}

\newcommand{\df}{\text{ d}}
\usepackage{hyperref}

\begin{document}

\markboth{Gonzalo J. Olmo and Diego Rubiera-Garcia}{}

%
\catchline{}{}{}{}{}
%

\title{The quantum, the geon, and the crystal}

\author{Gonzalo J. Olmo}

\address{ Departamento de F\'{i}sica Te\'{o}rica and
IFIC,  Universidad de Valencia-CSIC, Facultad de F\'{i}sica, C/ Dr. Moliner 50,
Burjassot-46100, Valencia, Spain.}
\address{Departamento de F\'isica, Universidade Federal da
Para\'\i ba, 58051-900 Jo\~ao Pessoa, Para\'\i ba, Brazil}

\author{Diego Rubiera-Garcia}

\address{Center for Field Theory and Particle Physics and Department of Physics, Fudan University, 220 Handan Road, 200433 Shanghai, China}

\maketitle

\begin{history}

\end{history}

\begin{abstract}
Effective geometries arising from a hypothetical discrete structure of space-time can play an important role in the understanding of the gravitational physics beyond General Relativity. To discuss this question, we make use of lessons from crystalline systems within solid state physics,  where the presence of defects in the discrete microstructure of the crystal determine the kind of effective geometry needed to properly describe the system in the macroscopic continuum limit. In this work we study metric-affine theories with non-metricity and torsion, which are the gravitational analog of crystalline structures with point defects and dislocations. We consider a crystal-motivated gravitational action and show the presence of topologically non-trivial structures (wormholes) supported by an electromagnetic field. Their existence has important implications for the quantum foam picture and the effective gravitational geometries. We discuss how the dialogue between solid state physics systems and modified gravitational theories can provide useful insights on both sides.
\end{abstract}



\section{Introduction}	

\subsection{The quantum}

The combination of theoretical models and new experimental results in the last decades has allowed to shape the so-called standard $\Lambda$CDM model of cosmology, based on Einstein's theory of general relativity (GR), and driven by a cold dark matter source, an inflationary process in the early Universe, and a tiny cosmological constant\cite{Weinberg}. Numerous observational data confirm the accuracy of this model for the explanation of a plethora of phenomena. These successes, however, cannot conceal other theoretical and experimental difficulties, which can be classified in two types:

\begin{itemize}

\item If GR holds true, then the presence of singularities deep inside black holes and in the early Universe is an unavoidable consequence resulting from a set of well established theorems\cite{Singularities1,Singularities2,Singularities3,Singularities4,Singularities5}. Though the very definition of space-time singularity is on itself a subject of physical and philosophical  debate\cite{Curiel2009},  it is clear that their existence poses considerable difficulties both at the conceptual and operational levels. It is widely believed that a consistent theory of quantum gravity (QG) should be free of such undesirable features\cite{QG1,QG2}.
\item The $\Lambda$CDM model requires the existence of large amounts of non-baryonic forms of matter (dark matter) of which we have no experimental evidence so far. Troubles with dark energy are even greater since we have no reliable picture of its nature, being only able to mimic its equation of state using exotic sources of matter/energy. The dark matter/energy framework seems too unnatural; an ad hoc construction designed to fit data sets not grounded on solid principles and lacking direct experimental support.

\end{itemize}

On these grounds, one could support the viewpoint that the underlying gravity theory is calling for modifications both in the ultraviolet and the infrared regimes\cite{Berti}. In fact,  why should high-energy new physics be restricted to the matter sector?  What kind of changes could be expected in the gravitational dynamics? What lessons/implications could be extracted from the high-energy regime to understand/interpret the gravitational dynamics at large scales?.

Though it is generally expected that the solution to those questions should be found within the framework of a quantum theory of gravity, the truth is that very little is known about it. The most reliable results in this sense are those coming from the quantization of fields in curved space-time\cite{QG2}, which seem to be in agreement with perturbative results of string theory\cite{strings} in the sense that higher-order curvature invariants are expected to play some role at increasing energies. Given that the number of observables can be limited, and the experimental precision is finite, at the end of the day it could happen that only a restricted sector of the full theory could be experimentally accessible. In this sense, effective approaches can be useful and, moreover, can provide relevant information towards the construction of the underlying theory\cite{Cembranos,Jimenez:2014fla}. This leads to the natural question of what kind of approaches could be suitable for an effective description of quantum gravity. In this paper we shall make use of three concepts - \emph{the quantum, the geon, and the crystal} - as useful ideas to explore this question.

\subsection{The geon}

The idealization of planets as bodies moving along geodesics of the space-time metric is one of the predictions of GR that has received the most thorough observational confirmation. However, though such an approximation is successful for macroscopic bodies, from a microscopic point of view elementary point particles appear as black holes, which entails singularities in the metric. Though it is assumed that such undesirable aspects will be avoided within QG, one could look instead for self-consistent descriptions of the notion of body within a purely classical gravitational treatment. The first realization of that idea was provided by Wheeler's geon\cite{Wheeler}. Geons were introduced as hypothetical entities describing self-gravitating electromagnetic bodies from the combination of GR and the Maxwell equations. In Wheeler's original formulation, this is achieved through balls of light, as a first step towards a singularity-free description of particles and gravity. With the seasoning of multiply-connected geometry, Misner and Wheeler showed\cite{Misner} that through a wormhole a sourceless electromagnetic field can create mass and charge (see Ref.\cite{Melvin} for a different realization of these concepts).

Within modern physics, the concepts of self-gravitating fields and non-Euclidean topologies play a role within a larger picture: that of space-time foam. In this picture of the quantum gravity world, the smoothness of Minkowski space-time would disappear at microscopic (Planckian) scales, where space-time is visualized as a highly non-trivial dynamical structure where not only geometry, but also topology, may fluctuate in all possible manners. In that microscopic regime the very concepts of distance and time become blurry and the challenge is to find new ideas and tools to deal with such a new scenario. It is worth pointing out that, given the smallness of the Planck scale, it is usually stated that the details of such an hypothetical microstructure should have a negligible effect on the macroscopic regime. To properly address this question, however, we need a suitable framework to describe how from a discrete, microscopic structure filled with topological defects, the macroscopic continuum space-time that we perceive could arise.

\subsection{The crystal}

To obtain new insights on this question it might be helpful to explore other domains of Physics. Indeed, though geometry made its appearance in Physics within  gravitation, there are other physical contexts where it has been shown to play an important role. This is the case, in particular, of crystalline structures such as Bravais crystals (see e.g.\cite{Kittel}). At the microscopic level they are defined by a discrete regular network of atoms. However, their continuum, macroscopic description is done through a procedure where the crystal is continuized in such a way that the lattice spacing is taken to zero while leaving the point mass unchanged. It turns out that differential geometry is the right tool to describe the resulting continuum effective geometry\cite{Kroner1a,Kroner1b}. This works as follows: at each point of the crystal, two or three lattice vectors define crystallographic directions. Moving along these vectors, one jumps from atom to atom in a natural step-counting process of measurement of distances. Affine connections allow to transport vectors and define geodesics in the crystal. The relevant result for our considerations here is that different kinds of crystals require different effective geometries. If the crystal is perfect, meaning that no defects are present, step-counting is possible, and the resulting effective geometry is Riemannian. However, all known ordered structures  contain defects, which can be of different types. It has been known for a long time that the continuum analog of dislocations (a line-like defect) is Cartan's torsion. Less known is the description for point defects. These can be of two types: intersticials, which are atoms that have leaven their equilibrium locations, and vacancies, corresponding to the holes left behind. Point defects prevent a simple step-counting, requiring the introduction of non-metricity, $\nabla^{\Gamma} g_{\mu\nu} \neq 0$\cite{Kroner2a,Kroner2b}. The effective geometry of a crystal with point-like and line-like defects is thus of metric-affine type with non-metricity and torsion\cite{Kroner3a,Kroner3b,Kroner3c,Kroner3d}.

Focusing on systems with point defects only, the step-counting procedure can be implemented by introducing a new (fictitious or idealized) non-defected crystal by the prescription that one just ignores point defects by filling vacancies and not counting intersticials at all. Measurements of length in this fictitious crystal can be carried out using a (fictitious or idealized) metric $h_{\mu\nu}$ compatible with the connection, $\nabla^{\Gamma} h_{\mu\nu}=0$, that transports vectors along the idealized crystallographic directions. A relation between $h_{\mu\nu}$ and the defect crystal metric $g_{\mu\nu}$ can be established once the density of defects is known. This procedure allows for a successful description of the defected crystal and its associated effective geometry.

Therefore, the idea that the space-time could have a microscopic structure pervaded by topological defects of different kinds (quantum foam) is somewhat analogous to the situation in real crystals. Consequently, useful lessons can be extracted from the fact that different geometric elements (non-metricity, torsion, ...) are necessary for a proper description of different kinds of defects. To implement these ideas, in this work we shall adopt the metric-affine (or Palatini) formalism\cite{olmo}, which relaxes any a priori relation between the metric and affine structures of the theory. We note that this is a foundational question of the idea of gravitation as a space-time phenomenon (since metric and affine connections carry very different meaning\cite{Zanelli}) as basic as the number of space-time dimensions, but which has received little attention so far. We point out that the answer to that question must be determined ultimately by experiment, not by tradition or convention.

In this framework the field equations are obtained as independent variations of the action with respect to metric and connection. In the case of GR, the connection turns out to be given by the Christoffel symbols of the metric (Riemannian geometry) due to the particular functional form of the Einstein-Hilbert Lagrangian, which makes the equation for the connection to simply express its compatibility with the metric. This is the same result that one obtains when the compatibility condition is imposed a priori (metric approach) and then the variation of the action with respect to the metric alone is performed (note that this also happens in the Lovelock family of Lagrangians, which includes GR as a particular case\cite{Lovelock1,Lovelock2,Lovelock3}). However, this is not necessarily so for actions extending GR, which is the case we are interested in. In what follows we shall thus consider a non-Riemannian framework with non-metricity and torsion.

\section{Metric-affine gravity with torsion}

Let us now specify the framework that shall be used throughout the paper. We consider a class of gravitational actions constructed as

\begin{equation} \label{eq:action}
S=\frac{1}{2\kappa^2} \int d^4x \sqrt{-g} f(R,Q) + S_m(\psi_m,g_{\mu\nu}),
\end{equation}
with the following conventions: $k^2$ is the gravitational constant in suitable units (in the case of GR, $\kappa^2=8\pi G$), $g$ is the determinant of the space-time metric $g_{\mu\nu}$, which is a priori independent of the affine connection $\Gamma_{\mu\nu}^{\lambda}$ (metric-affine or Palatini approach\cite{olmo}), the Lagrangian $f(R,Q)$ is a given function of the objects $R=g_{\mu\nu}R^{\mu\nu}$ and $Q=R_{\mu\nu}R^{\mu\nu}$ where $R_{\mu\nu}(\Gamma)$ is the Ricci tensor, which is obtained from the Riemann tensor as $R_{\mu\nu}\equiv {R^\alpha}_{\mu\alpha\nu}$, and the matter action $S_m$ is assumed to depend on the metric only, while $\psi_m$ denote collectively the matter fields.

We further assume that the Lagrangian can be expressed as $f(R,Q_S,Q_A)$, where $Q_S=R_{(\mu\nu)}R^{(\mu\nu)}$ and $Q_A=R_{[\mu\nu]}R^{[\mu\nu]}$ are constructed with the symmetric and antisymmetric parts of the Ricci tensor, respectively. Variation of the action (\ref{eq:action}) with respect to the metric $g_{\mu\nu}$ yields the corresponding field equations

\begin{equation} \label{eq:metric}
\kappa^2 T_{\mu\nu}=\frac{\partial f}{\partial g^{\mu\nu}} -\frac{f}{2}g_{\mu\nu},
\end{equation}
where we have defined $f_R \equiv df/dR$ and $T_{\mu\nu}\equiv -\frac{2}{\sqrt{-g}}\frac{\delta S_m}{\delta g^{\mu\nu}}$ is the energy-momentum tensor of the matter. On the other hand, the field equations resulting from the variation with respect to the connection $\Gamma_{\mu\nu}^{\lambda}$ can be conveniently expressed into their symmetric and antisymmetric parts as\cite{Torsion}

\begin{eqnarray}
\frac{1}{\sqrt{-g}}\nabla_{\alpha}[\sqrt{-g}M^{(\beta \nu)}]&=&{\Lambda^{(+)} }_{\alpha}^{\beta \nu \kappa \rho} M_{[\kappa \rho]} \label{eq:con1a} \\
\frac{1}{\sqrt{-g}}\nabla_{\alpha}[\sqrt{-g}M^{[\beta \nu]}]&=&{\Lambda^{(-)} }_{\alpha}^{\beta \nu \kappa \rho} M_{(\kappa \rho)}, \label{eq:con1b}
\end{eqnarray}
where $M^{(\beta \nu)}=f_R g^{\beta \nu} + 2f_{Q_S} R^{(\beta \nu)}(\Gamma)$ and $M^{[\beta \nu]}=2f_{Q_A} R^{[\beta \nu]}(\Gamma)$. The coupling between these two parts is mediated by the objects
\begin{equation}
{\Lambda^{(\pm)} }_{\alpha}^{\beta \nu \kappa \rho}=\left[\widetilde{S}_{\alpha \lambda}^{\nu} g^{\beta \kappa} \pm \widetilde{S}_{\alpha \lambda}^{\beta} g^{\nu \kappa} \right] g^{\lambda \rho}.
\end{equation}
which are constructed with the torsion $S_{\mu\nu}^{\lambda} \equiv \Gamma_{\mu\nu}^{\lambda} - \Gamma_{\nu\mu}^{\lambda}$ through the traceless tensor $\widetilde{S}_{\mu\nu}^{\lambda}=S_{\mu\nu}^{\lambda}+\frac{1}{3} (\delta_{\nu}^{\lambda}S_{\mu}-\delta_{\mu}^{\lambda} S_{\nu})$, where $S_{\mu}=S_{\lambda \mu}^{\lambda}$.

The system of equations (\ref{eq:metric}), (\ref{eq:con1a}) and (\ref{eq:con1b}) describes an effective non-Riemannian geometry with non-metricity, $\nabla^{\Gamma}g_{\mu\nu} \neq 0$, and torsion, $S_{\mu\nu}^{\lambda} \neq 0$. This theory, therefore, contains all the elements necessary to describe a continuous system  which could correspond to a crystalline structure with point defects (vacancies/intersticials) and dislocations. The search for solutions to these equations, however, is a highly non-trivial task, which seems to be possible only in simplified scenarios, at least analytically. For example, when torsion vanishes (which is the analog of dislocations being neglected), $S_{\mu\nu}^{\lambda} = 0$, the field equations for the metric become

\begin{equation}
f_R R_{(\mu\nu)}-\frac{f}{2}g_{\mu\nu} + 2g^{\alpha\beta} f_{Q_S}R_{(\mu\alpha)}R_{(\nu\beta)}   = \kappa^2 T_{\mu\nu} - 2g^{\alpha\beta} f_{Q_A}R_{[\mu\alpha]}R_{[\nu\beta]} \ , \label{eq:met1sPost}
\end{equation}
If we further assume the Lagrangian density to be written as $f(R,Q_S,Q_A)=\tilde{f}(R,Q_S)+\hat{f}(Q_A)$, then we can move the piece $-\frac{\hat{f}}{2}g_{\mu\nu}$ to the right-hand-side, and interpret the extra terms there as a kind of effective non-linear electrodynamics energy-momentum tensor.

Taking appropriate traces on these equations yields the result $R=R({T_\mu}^{\nu})$ and $Q_S=Q_S({T_\mu}^{\nu})$, which means that these geometrical invariants only depend on the matter sources, ${T_\mu}^{\nu}$. On the other hand, in this torsionless scenario the equations for the connection (\ref{eq:con1a}) and (\ref{eq:con1b}) decouple, which largely simplifies their resolution. If we further impose $R_{[\mu\nu]}=0$ (so $Q_A=0$) the only non-trivial equations are those of (\ref{eq:con1a}), which can be conveniently expressed as

\begin{equation} \label{eq:h}
\nabla_{\alpha}^{\Gamma}[\sqrt{-h}h^{\beta \nu}]=0,
\end{equation}
where $h_{\mu\nu}$ is a new rank-two tensor. Note that this equation simply expresses the compatibility between the independent connection $\Gamma_{\mu\nu}^{\lambda}$ and the metric $h_{\mu\nu}$ which, therefore, has associated a Riemannian geometry.  One can show that $h_{\mu\nu}$ is related to $g_{\mu\nu}$ as
\begin{equation}\label{eq:hmn}
h^{\mu\nu}=\frac{g^{\mu\alpha}{\Sigma_\alpha}^\nu}{\sqrt{\det \Sigma}}  \ ; \   h_{\mu\nu}\equiv (\sqrt{\det \Sigma}) {[\Sigma^{-1}]_\mu}^\alpha g_{\alpha\nu} \ ,
\end{equation}
where the matrix ${\Sigma_\alpha}^\nu$ is defined as
\begin{equation}\label{eq:Sigma}
{\Sigma_\alpha}^\nu=f_R  {\delta_\alpha}^\nu +2f_{Q_S}{P_\alpha}^\nu \ ,
\end{equation}
with ${P_\alpha}^\nu\equiv R_{\mu\alpha} g^{\alpha \nu}$.
Given that ${\Sigma_\mu}^\nu$ is a function of ${T_\mu}^{\nu}$, we see that the relation between $h_{\mu\nu}$ and $g_{\mu\nu}$ is analogous to that found in defected crystals, but with the density of defects replaced by the stress-energy density.  This suggests that ${T_\mu}^{\nu}$ can be seen as a sort of coarse-grained description of phenomena occurring at the microscopic level.

In terms of $h_{\mu\nu}$, the metric field equations (\ref{eq:met1sPost}) can be written as

\begin{equation}\label{eq:Rmn0Post}
{R_\mu}^\nu (h)= \frac{1}{\sqrt{\det \Sigma}} \left[\frac{f}{2}{\delta_\mu}^\nu  + \kappa^2 {T_\mu}^\nu \right]  \ .
\end{equation}
Given that $f(R,Q_S)$ can always be expressed as a function of the energy-momentum tensor $T_{\mu\nu}$, these equations can be seen as a system of second-order Einstein-like equations with a modified stress-energy source on the right-hand-side. On the other hand, the algebraic transformations (\ref{eq:hmn}), which are governed by the matrix ${\Sigma_\mu}^\nu$ in Eq.(\ref{eq:Rmn0Post}) (which only depends on the energy-momentum sources), generate a non-Riemannian structure for $g_{\mu\nu}$ with second-order field equations as well. Once a gravity Lagrangian $f(R,Q_S)$ is specified, this procedure provides a full solution to a given problem. In absence of defects/matter, ${T_\mu}^\nu=0$, one finds $h_{\mu\nu}=g_{\mu\nu}$ and the equations (\ref{eq:Rmn0Post}) boil down to those of GR plus a cosmological constant. This clearly shows that the existence of an effective non-Riemannian geometry is due to the presence of defects and their interaction with the gravitational sector of the theory. That the vacuum equations of the theory are those of GR also implies that no extra propagating degrees of freedom arise, and the theory is free of ghost-like instabilities. This provides a consistent framework to study the implications of the existence of point defects in the microstructure for the effective geometry at the macroscopic level. In the next section we shall employ a suitable gravity Lagrangian for this purpose.

\subsection{Born-Infeld gravity and point defects}

The question we face now is: what kind of Palatini Lagrangian could properly account for the presence of topological defects in the microstructure of space-time?. To obtain insights on this question, we look again for answers within the physics of solid state physics and the description of point defects in crystalline structures. Following Katanaev and Volovich\cite{Katanaev}, the mass of a point defect can be estimated as the difference of volume between the defected crystal and the perfect crystal that one would have if no defects were present. In turns out that there is a gravitational action that can be expressed in that form, namely

\begin{equation} \label{eq:BIaction}
S_{BI}= \frac{1}{\kappa^2 \epsilon} \int d^4x (\sqrt{-q} - \sqrt{-g})
\end{equation}
where $\epsilon$ is a small parameter with dimensions of length squared and we have introduced another metric $q_{\mu\nu}=g_{\mu\nu}+\epsilon R_{\mu\nu}$. This Lagrangian is the natural extension, to the gravitational context, of Born and Infeld non-linear theory of electrodynamics\cite{BI-matter} through the replacement of the field strength tensor $F_{\mu\nu}$ by the Ricci tensor $R_{\mu\nu}$ and, consequently, is dubbed as Born-Infeld (BI) gravity\cite{BI-gravity1,BI-gravity2}. In the same way as Born-Infeld electrodynamics cured the problem of the divergence of the self-energy of the electron in classical electrodynamics, Born-Infeld gravity could help to solve the problem with singularities in GR. Its physical meaning is the comparison of volume elements between the ``perfect" metric $q_{\mu\nu}$ and the ``defected" metric $g_{\mu\nu}$. This follows from the fact that the metric $q_{\mu\nu}$ is compatible with the connection, $\nabla^{\Gamma}_{\alpha}[\sqrt{-q} q^{\mu\nu}]=0$, which implies that the independent connection $\Gamma^{\lambda}_{\mu\nu}$ is given by the Christoffel symbols of $q_{\mu\nu}$, playing the same role as $h_{\mu\nu}$ in Palatini $f(R,Q)$ theories above.

Expanding the action (\ref{eq:BIaction}) in series of $\epsilon$ one obtains

\begin{equation} \label{eq:quadratic}
S \approx \frac{1}{2\kappa^2}\int  d^4x \sqrt{-g}  \left[R -\frac{\epsilon}{2} \left(-\frac{R^2}{2} + R_{\mu\nu}R^{\mu\nu}\right)+\ldots\right] \ , \nonumber
\end{equation}
whose leading order term corresponds to the Einstein-Hilbert action of GR, where we have $\sqrt{-q}=\sqrt{-g}$ (which means absence of point defects) and an effective Riemannian geometry describing a non-defected crystal.  When corrections in $\epsilon$ are taken into account, one gets a non-Riemannian theory of the form $f(R,Q)$, describing a system with point defects due to the presence of non-metricity, $\nabla_{\alpha}^{\Gamma} (\sqrt{-g} g^{\mu\nu}) \neq 0$. The relation between $q_{\mu\nu}$ and $g_{\mu\nu}$ is formally the same as in the $f(R,Q)$ theories [see Eqs.(\ref{eq:hmn}) and (\ref{eq:Sigma})], which means that also in this case the existence of non-metricity is caused by the energy-density of the matter fields.

To complete this framework we note that the field equations for this theory in terms of $q_{\mu\nu}$ can also be cast in a formally similar way as those of Eqs. (\ref{eq:Rmn0Post}), namely

\begin{equation}\label{eq:BIfield}
{R_\mu}^\nu (q)= \frac{\kappa^2}{\vert \hat{\Sigma} \vert^{1/2}} \left[L_G {\delta_\mu}^\nu +  {T_\mu}^\nu \right]  \ ,
\end{equation}
where $L_G$ is the BI gravity Lagrangian, which can be expressed as

\begin{equation}
L_G=\frac{\vert \hat{\Sigma} \vert^{1/2} -\lambda}{\epsilon \kappa^2}
\end{equation}
In vacuum, the same considerations regarding the vacuum structure of the field equations of the case of $f(R,Q)$ gravities also apply here.

\subsection{Palatini geons}

The analogy put forward above between the distribution of point defects in crystals and the density of energy-momentum in theories of gravity with non-metricity suggests that these theories can be used as a means to explore effective descriptions of the quantum gravitational world. To go farther in this direction, we will now show that geonic structures \`{a} la Wheeler can naturally arise in these scenarios. To proceed, we consider a simplified scenario corresponding to a static, spherically symmetric electromagnetic field (without sources), with a single non-vanishing component $F_{tr} \equiv E(r)$. The associated solution of the combined system of BI gravity (and also of the quadratic Lagrangian (\ref{eq:quadratic})) with a Maxwell field has been obtained in exact analytical form and we bring here the main result. This can be expressed, in Eddington-Finkelstein coordinates, as \cite{or12a,or12b,or12c,ora}
\begin{equation}\label{eq:ds2_EF}
ds^2=-A(x)dv^2+\frac{2}{\sigma_+}dvdx+r^2(x)d\Omega^2 \ ,
\end{equation}
where the metric function takes the form
\begin{equation}\label{eq:A}
A(x)= \frac{1}{\sigma_+}\left[1-\frac{r_S}{ r  }\frac{(1+\delta_1 G(r))}{\sigma_-^{1/2}}\right]
\end{equation}
with the definitions

\begin{equation}
\delta_1= \frac{1}{2r_S}\sqrt{\frac{r_q^3}{l_P}} \  ; \ \sigma_\pm=1\pm \frac{r_c^4}{r^4(x)} \ ; \ \frac{dr}{dx} = \frac{\sigma_{-}^{1/2}}{\sigma_+} \label{eq:r(x)} \ ,
\end{equation}
where we have redefined the parameter $\epsilon$ as $\epsilon=-2al_P^2$ (since we assume $\epsilon<0$), with $a>0$ a dimensionless constant, introduced a constant $r_c$, defined as $r_c=\sqrt{l_P r_q}$, where $l_P=\sqrt{\hbar G/c^3}$ is Planck's length and $r_q^2=2G_N q^2$ is a length scale associated to the electric charge, and $r_S=2M_0$ is Schwarzschild radius. The function $G(z)$, with $z=r/r_c$, can be written as an infinite power series expansion of the form
\begin{equation}
G(z)=-\frac{1}{\delta_c}+\frac{1}{2}\sqrt{z^4-1}\left[f_{3/4}(z)+f_{7/4}(z)\right] \ ,
\end{equation}
where $f_\lambda(z)={_2}F_1 [\frac{1}{2},\lambda,\frac{3}{2},1-z^4]$ is a hypergeometric function, and $\delta_c\approx 0.572069$ is a constant.

\begin{figure}[h]
\begin{center}
\includegraphics[width=0.5\textwidth]{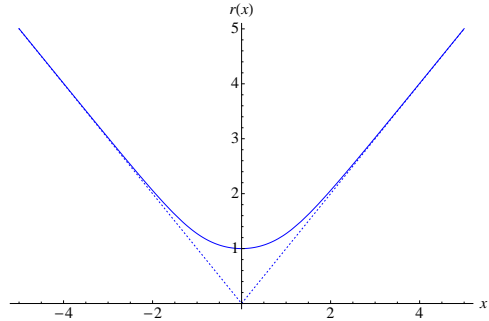}
\caption{The function $r(x)$ (solid curve), defined in (\ref{eq:r(x)}), as a function of the radial coordinate $x$  in units of the scale $r_c$. The dotted lines represent the function $|x|$ (GR case), to which $r(x)$ approaches as $x\rightarrow \infty$. At $x=0$ we have $z=1$ and a bounce in the radial coordinate occurs, revealing the existence of a wormhole structure. }\label{fig:1}
\end{center}
\end{figure}

The above space-time metric has a number of interesting properties. It represents a generalization of the standard Reissner-Nordstr\"om (RN) solution of GR, which is recovered for $z\gg 1$, as can be easily checked from the corresponding expansion ($G(z)\approx -1/z$, $\sigma_\pm \approx 1$, $r^2(x)\approx x^2$)
\begin{equation} \label{eq:RNsolution}
A(x)\approx 1-\frac{r_S}{ r  }+\frac{r_q^2}{2r^2} \ .
\end{equation}
According to this, for astrophysical size black holes, these objects look like the RN solution in terms of the location of their (external) event horizon. However, the innermost structure of this space-time is completely different from its GR counterpart. The most relevant point is related to the behaviour of the radial function $r^2(x)$, which follows from integration of Eq.(\ref{eq:r(x)}) as

\begin{equation}
r^2(x)= \frac{x^2+\sqrt{x^4+4r_c^4}}{2}
\end{equation}
As depicted in Fig.\ref{fig:1}, the function $r^2(x)$, which measures the area of the two-spheres, attains a minimum at $x=0$, before bouncing off. The resulting structure is reminiscent of a wormhole geometry, with its throat located at $x=0$ ($z=1$), though some additional considerations are necessary before we can conclude that it is indeed the case.
To proceed, note that the solutions that we have obtained correspond to a sourceless electromagnetic field. In the standard electrically charged RN black hole solutions of GR, it is natural to wonder where the electric charge generating the field is placed. This leads to a well known inconsistency. In fact, it is impossible for a point-like charge placed at the black hole center to account for the whole mass and charge of space-time and, at the same time, be a solution of Einstein's equations everywhere\cite{Ortin}. In our case, the electric charge follows from the integral of the lines of flux flowing through a closed $2-$surface that encloses the $r=r_c$ region as
\begin{equation}
\Phi= \int_{S}*F =4\pi q \ .
\end{equation}
Since there are no sources in our setup, which consists just on a free electric field, this nonzero electric flux defining the (conserved) charge $q$ must have a topological origin. The hypersurface $x=0$, where $r=r_c$ reaches its absolute minimum, represents a hole in the topology through which the electric field lines flow. As a result, no point-like charges are necessary to explain the presence of $q$. Now, since the wormhole connects two spaces (or, equivalently, has two sides) a similar computation can be carried out on the other side of the hole. If the orientation of the $2-$surface is taken in the direction of growth of the $2-$spheres, then the resulting flux would give $-4\pi q$. This means that the electric field lines enter through one of the sides of the wormhole and exit through the other, defining a negative charge on one side and a positive charge on the other. If in this space-time there were $2-$surfaces able to enclose the two sides of the wormhole at the same time, then the net flux through them would be zero.

Remarkably, the electric field of these wormholes has an interesting universal property. Given that the area of the throat grows linearly with the charge, $A_{WH}\propto r_c^2=l_p r_q\sim |q|$, the amplitude of the electric field, defined as the ratio of the flux by surface area, at the throat becomes
\begin{equation}
\frac{\Phi}{4\pi r_c^2} = \sqrt{\frac{c^7}{2\hbar G^2}} \ ,
\end{equation}
which is  independent of the specific amounts of mass and charge present in the system. Moreover, this quantity only depends on the fundamental constants associated with relativistic ($c$) quantum ($\hbar$) gravity ($G$). Thus, the size of the wormhole increases with the intensity of the electric flux in such a way that the amplitude of the electric field at the throat stays constant. This universal behavior  is valid for arbitrary charge-to-mass ratio $\delta_1$, putting forward that all configurations have identical topological properties.  This is relevant because configurations with a certain value of $\delta_1$ are completely free of curvature divergences, whereas for arbitrary $\delta_1$ curvature scalars typically diverge at the throat. The universality of the electric field at the throat, therefore, suggests that the pathological behavior typically attributed to curvature divergences should be reconsidered with more detail in this scenario.

Theorems on existence of singularities rely on three assumptions: formation of trapped surfaces, fulfillment of the energy conditions, and existence of global hyperbolicity\cite{Singularities1,Singularities2,Singularities3,Singularities4}. Though this holds true for GR, it does not necessarily apply to extensions of GR like the ones considered here. Moreover, though the blow up of curvature invariants is usually employed as a way to detect the existence of space-time singularities, it has been well understood for a long time\cite{Geroch} that a more reliable criterion to determine whether a space-time is singular or not, is provided by the notion of geodesic completeness. According to this criterion, if {\it all} time-like and null geodesics can be extended to arbitrarily large values of the affine parameter, the space-time is said to be geodesically complete (nonsingular), while if there exist curves terminating on a finite value of the affine parameter (and no further extension is possible) then a space-time singularity arises. Since in our case, curves with $\delta_1 \neq \delta_c$, which present curvature divergences, can be arbitrarily close to the case with bounded curvature scalars  ($\delta_1=\delta_c$), the  ability of curvature invariants to determine the existence of space-time singularities is put into question. One should thus study the behavior of geodesic curves, described by the equation
\begin{equation}
  \frac{\df^2  x^\mu}{\df  \lambda^2}+\Gamma^\mu_{\alpha \beta} \frac{\df x^\alpha}{\df \lambda}\frac{\df x^\beta}{\df \lambda} = 0 \ , \label{eqgeo}
 \end{equation}
where $\lambda$ is the affine parameter, in this geonic backgrounds to determine if this space-time is geodesically complete or not. Since the connection is not directly coupled to the matter sector in the action (\ref{eq:action}), the natural set of geodesics to explore are those associated to the causal structure of $g_{\mu\nu}$. In addition, following\cite{Tipler} one should also consider the effect of the presence of a curvature divergences on the forces acting upon extended bodies, and whether this implies that such bodies are destroyed (or not) when crossing the $x=0$ surface.

As advanced above, the Born-Infeld gravity action (\ref{eq:BIaction}) can be seen as related to the mass of a point defect. To further strengthen this view, we can evaluate this action on the solutions just obtained. This yields the result
\begin{equation}
S=2M_0 \frac{\delta_1}{\delta_c} \int c dt \ ,
\end{equation}
which is the action of a point-like particle with total energy $\epsilon =M_0 \delta_1/\delta_c$, where $M_0$ is the total mass of space-time! (note that the factor two in this result comes from the need to integrate on both sides of the wormhole).  The case $\delta_1= \delta_c$, where curvature divergences are absent, can somewhat be understood as the absence of forces acting upon the geon and leads to $\varepsilon=M_0$, while when $\delta_1  \neq \delta_c$ a disturbance in the energy of the geon occurs, which is corrected by a factor $\delta_1/\delta_c$. Geons can be interpreted as a kind of gravitational soliton: the closest thing to the notion of a classical body in presence of gravity. They can generate their own charge and mass and posses a topological characterization. We have considered static solutions but, in order to strengthen the analogy with point defects in crystalline structures one should also study their dynamical behavior. This is due to the fact that, as we have already mentioned, point defects in crystals actually have dynamics: they can be created and annihilated by external forces, heat or intense fields, are able to move through the crystal and to interact and recombine with the same or different kinds of defects. Such a behaviour resembles the creation/annihilation of point-like objects with non-trivial topology like the ones considered here and suggests an extension of the static solution to include dynamic behaviour. As a first step to get further into this issue, let us point out that the static geonic wormholes above can be dynamically generated out of Minkowski space. This has been explicitly shown in the case of simple models with a flux of particles carrying energy and charge\cite{lmor}. In this sense, if one starts with Minkowski space, the process of absorption of mass and charge can be modelled in terms of a charged Vaidya-like solution\cite{BVa,BVb}. When the flux ceases one recovers the static solutions presented before. The possibility of forming wormholes dynamically supports the view of a space-time foam generated by quantum fluctuations at microscopic scales. This should be investigated in more detail, for example within scenarios of formation of pairs of microscopic wormholes in highly magnetized scenarios\cite{cylindrically1,cylindrically2,cylindrically3,cylindrically4,cylindrically5}. Further research should determine if virtual pair production could be seen as the formation of wormholes. On the other hand, one could also investigate whether the standard dynamical process of realistic gravitational collapse of macroscopic bodies could be modified in the last stages of the collapse due to quantum gravity effects\cite{Bambi1,Bambi2,Bambi3}, in such a way that the GR point-like singularity is truly replaced by a wormhole structure.

\section{Further comments and outlook}

Given that all ordered crystalline systems in Nature present defects of different kinds and that such systems require metric-affine geometries for their proper description,  it is tempting to conjecture that Riemannian structures other than GR are somehow unstable or unsuitable for the description of non-idealized systems. This is supported, at least in part, by the generic existence of ghost-like instabilities in high-curvature extensions of Einstein's theory in the usual Riemannian approach, whereas in the metric-affine extensions shown here the dynamical equations are always second-order and smoothly converge to the GR equations in the vacuum (undefected) limit. From this viewpoint, GR appears as a natural attractor with robust dynamics.

On the other hand, the fact that we perceive an approximately Riemannian geometry (as orbital motions confirm) could be related to the fact that the density of defects at the scales measured so far is relatively low. At higher energies or  in strong curvature scenarios, the role of defects could be more relevant, as the modification of the Reissner-Nordstr\"{o}m geometry presented here suggests. The modifications induced by our model are intrinsically non-perturbative, as they involve topological effects.

To conclude, we note that the distinction between the discrete microscopic structure and the continuum macroscopic description is, to some extent, a matter of choice of the theoretician, a decision taken in order to match the modeling of the physical system with the observed facts. In the last decades, a discussion has grown of whether Physics should be seen instead as a multi-scale description, in which the phenomena occurring at \textit{any} chosen scale represent a sort of cumulative effects of the underlying events at some smaller scale \cite{Accardi}. For crystalline structures, even if the effective geometry at a given scale, where the system can be treated as a continuum, might seem to be Riemannian, the existence of point defects in the microscale can have a significant impact on the existence and properties of collective phenomena, such as plasticity or viscosity. How can these lessons be useful for the physics of gravitational phenomena and their relation with the existence of an hypothetical microscopic structure? At this point of our research we can only conjecture that the energy-momentum tensor should be seen as a coarse-grained description of the microscopic structure, and the effective geometries as an emergent phenomenon. Moreover, since different defects in different crystals are governed by different laws, this raises the question if different effective geometries could arise at different scales in space-time. Note that galaxies and their arrangements in filaments and other complex structures could be visualized as the microscopic defects of a larger scale effective geometry with properties and dynamics very different from those corresponding to smaller scales. In addition to deal with the problem of singularities, the reconsideration of foundational issues of GR and gravitation as geometric phenomena could help  identify new avenues to address  the important questions posed by the  dark matter/energy paradigm from a different perspective.

\section*{Acknowledgments}
G.J.O. is supported by the Spanish grant FIS2011-29813-C02-02, the Consolider Program CPANPHY-1205388, a Ramon y Cajal contract, and  the i-LINK0780 grant of CSIC. D.R.-G. is supported by the NSFC (Chinese agency) grant Nos. 11305038 and 11450110403, the Shanghai Municipal Education Commission grant for Innovative Programs No. 14ZZ001, the Thousand Young Talents Program, and Fudan University. The authors also acknowledge support from CNPq (Brazilian agency) grant No. 301137/2014-5.

\end{document}